\documentclass[11pt,twoside]{article}
\usepackage{asp2004}
\usepackage{psfig}
\usepackage{epsf}
\usepackage{graphics}
\usepackage{lscape}
\def\edcomment#1{\iffalse\marginpar{\raggedright\sl#1\/}\else\relax\fi}
\reversemarginpar

\begin{document}
\title{Multidimensional analysis of X-ray variability of AGN}
\author{A. V. Halevin}
\affil{Department of Astronomy, Odessa National University, T.G.Shevchenko park, 65014, Odessa, Ukraine
}

\begin{abstract}
In this work we analyzed X-ray light curves of active galactic nucleus NGC 4051 obtained 
using Advanced  CCD Imaging Spectrometer of Chandra satellite. 
Taking into account mainly flaring behaviour of AGNs we have used wavelet analysis for
searching of short time lived events on light curves.
\end{abstract}
\thispagestyle{plain}

\section{Introduction}

Short time-scale X-ray variability from dozens of seconds to hours  in
AGNs is explained as the result of different kinds of processes  which
happen close to the central engine. The possible origins are  changes of
the accretion rate, flares in accretion disk, motion of  hot spots
around the black hole and, for longer time scales,  motion of hydrogen
clouds which obscure the central source, etc. 

Here we present results of investigations of short time-scale 
variability of active galactic nuclei using as example Chandra 
observations of bright Seyfert type galaxy NGC~4051. All archive data sets (obsid 829, 2983 and 3144) were 
processed using CIAO 2.3 package.

\section{ANALYSIS OF VARIABILITY OF NGC 4051}

Seyfert type galaxy NGC~4051 is very bright object with rapid and dramatic changes of flux.

Chandra observations of NGC~4051 were made during AO1, AO2 and AO3 (see
Table 1).

\begin{table}[b]
\caption{Observation Log.} \label{tab1} 
\begin{tabular}{rlrc}
\hline Obs.ID&Date&Exp.ks,&Mode\\ 
\hline 
829&00-03-24/25&80.8&ACIS\\ 
2148&01-02-06&50.5&ACIS\\ 
3144&01-12-31/02-01-01&91.7&HRC\\
\hline 
\end{tabular}
\end{table}

In order to investigate the character of  variability, we used the
wavelet method \Citep{and1999}, which is excellent for  detection of
incoherent or weakly coherent variations. Many  papers are devoted to
the application of the wavelet method to  periodic or multiperiodic
processes \Citep{fr1998}. 

Long-term variations of NGC 4051 have  been smoothed by using the method
of the running parabola  \Citep{and1997}. The optimal value of the
filter half--width  $\Delta t=0.217$~d has been determined from
maximization of the  ``signal-to-noise" ratio. To avoid apparent effects
of  low--frequency trends on the test--functions at high frequencies, 
the original data have been detrended, i.e. the running parabola  fit
was subtracted from the observations. For these detrended time  series,
the test functions have been computed using the code  described by 
\Citet{and1994}. For visualization, we used the  Weighted Wavelet
Z-transform (WWZ) test-function \Citep{fos1996},  which is characterized
as having the best contrast among other  test functions. 

\begin{figure}[!b]
\vspace{0cm} \hspace{0cm} 
\resizebox{\hsize}{!}{\includegraphics{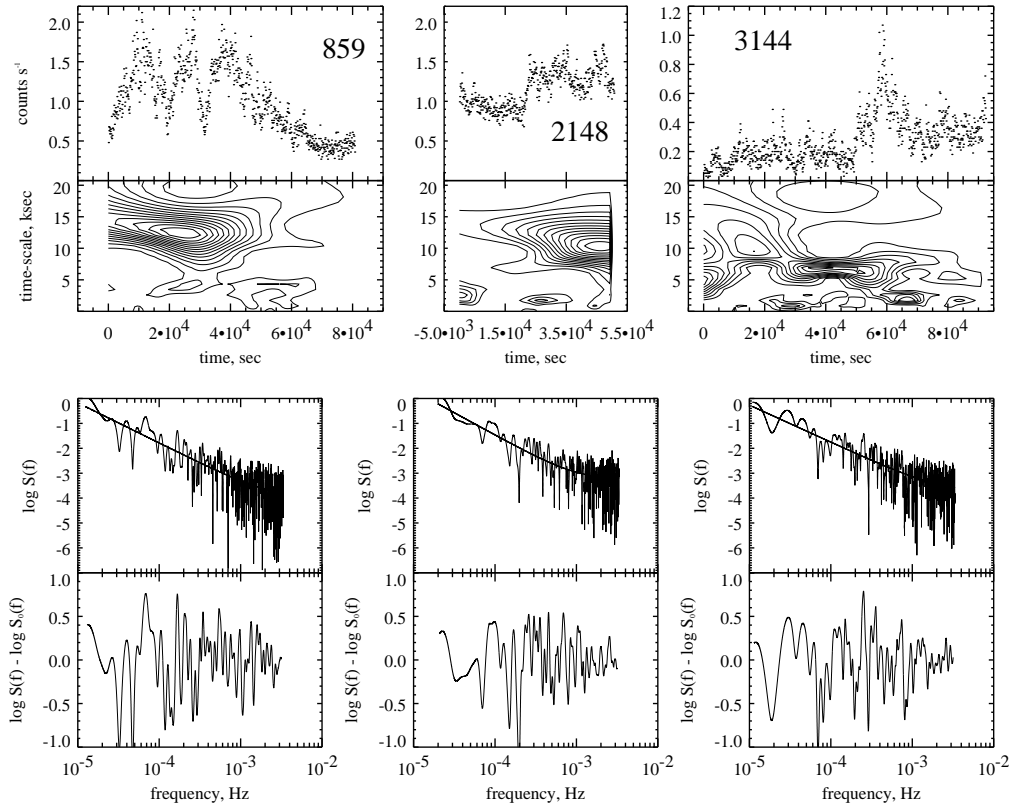}} \caption{Light curves, wavelet maps and power spectra
 for detrended light curves of NGC~4051.} \label{fig1} 
\vspace{0cm} 
\end{figure}

On Fig.1 one can see resulting wavelet maps, power spectra with red
noise trend and detrended power spectra for all data sets. Power spectra
for different runs show prominent peaks (Fig.2), but they never found
again for other runs. However, the same result for time-scale of about 4
ksec have been found by previous investigators \Citep[see][]{pl1995}. As
a result we can conclude incoherent ore only weakly coherent behaviour
of variability of NGC 4051.

To study typical time-scales of variability of NGC 4051 we plot the
distributions of time-scales derived from wavelet maps for different
runs (Fig.3). All plots show two different groups of peaks with
time-scale of approximately 5 and 15 ksec. Additional powerful peak (9
ksec) on distribution for the run 3144 corresponds to oscillations which
were observed during the large flare on light curve.

\begin{center}
\begin{figure}[!b]
\vspace{0.7cm} \hspace{2cm} 
\resizebox{8cm}{!}{
\includegraphics{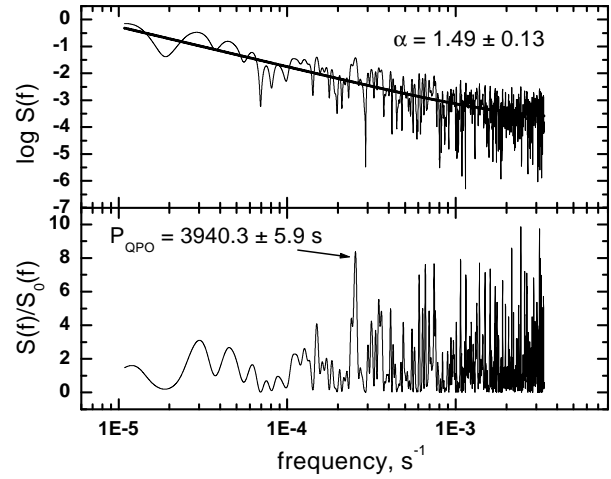}\\
} \caption{Initial and detrended power spectrum for run 3144 with peak around 3940 sec.} \label{fig2} 
\end{figure}

\begin{figure}[!b]
\vspace{0.7cm} \hspace{2cm} 
\resizebox{8cm}{!}{
\includegraphics{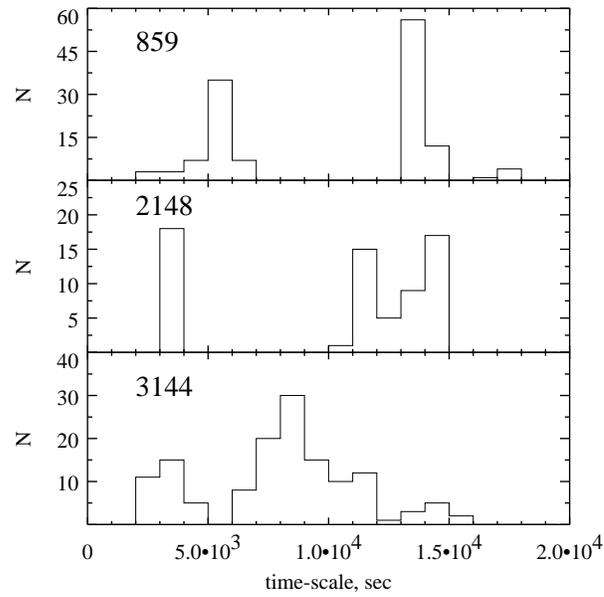}\\
} \caption{The distribution of the variability time-scales derived from wavelet maps
.} \label{fig3} 
\vspace{0cm} 
\end{figure}
\end{center}

On the next figure (Fig.4) we show evolution of time scales and
amplitudes for the most prominent oscillations detected with wavelet
analysis. One can see that mostly long lived oscillations have life time
approximately 3-4 times longer than their periods. This fact allow us
assume non flare nature of these events.

\begin{figure}[!t]
\vspace{0.7cm} \hspace{0cm} 
\resizebox{\hsize}{!}{
\includegraphics{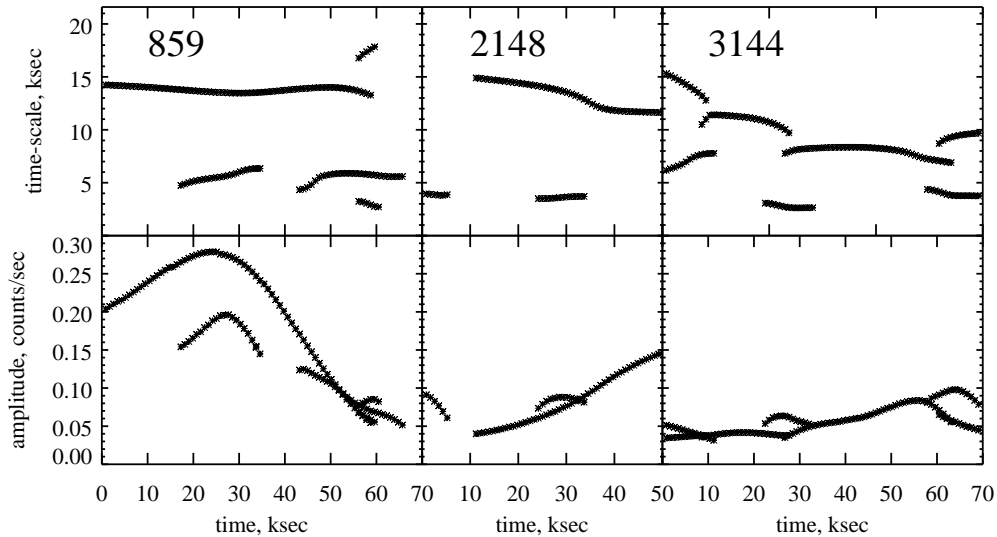}\\
} \caption{Quasi-coherent events on light curves of NGC 4051.} \label{fig4} 
\vspace{0cm} 
\end{figure}

\section{CONCLUSIONS}

Using results of wavelet analysis we declare absence of ``true"
quasi-periodical oscillations in NGC 4051 with coherence time-scale
longer than 4 periods. Detected QPO-like events with coherence
time-scale shorter than 3 periods must be explained as composition of
different flares on light curves. However, several events, observed
during 3-4 cycles, could be generated by some quasi-periodical processes
in the vicinity of central black hole.


\begin{thebibliography}{}

\bibitem[Andronov(1994)]{and1994}
Andronov, I.L. 1994,
Odessa Astron. Publ., 7, 49-54.

\bibitem[Andronov(1997)]{and1997}
Andronov, I.L. 1997,
\aaps, 125, 207.

\bibitem[Andronov(1999)]{and1999}
Andronov, I.L. 1999,
in: "Self-Similar Systems", eds. V.B. Priezzhev, V.P. Spiridonov,
 Dubna, JINR, 1999, 57-70.

\bibitem[Foster(1996)]{fos1996}
Foster, G. 1996,
\aj, 112, 1709

\bibitem[Fritz \& Bruch(1998)]{fr1998}
Fritz, T. \& Bruch, A. 1998,
\aap, 332, 586

\bibitem[Papadakis \& Lawrence(1995)]{pl1995}
Papadakis, I.E. \& Lawrence, A. 1995,
\mnras, 272, 161


\end{thebibliography}
\end{document}